\documentstyle[aps,psfig,twocolumn]{revtex}
\begin{document}
\draft
\flushbottom
\twocolumn[
\hsize\textwidth\columnwidth\hsize\csname @twocolumnfalse\endcsname

\title{Two-phase behavior in strained thin films of hole-doped manganites}
\author{Amlan Biswas, M. Rajeswari, R. C. Srivastava, 
Y. H. Li, T. Venkatesan,
and R. L. Greene}
\address{Center for Superconductivity Research, University of Maryland,
College Park, MD-20742}
\author{A. J. Millis}
\address{Center for Materials Theory,
Department of Physics and Astronomy, 
Rutgers University, Piscataway, NJ-08854}
\date{\today}
\maketitle
\tightenlines
\widetext
\advance\leftskip by 57pt
\advance\rightskip by 57pt

\begin{abstract}
We present a study of the effect of biaxial strain on the electrical and magnetic properties of thin films
of manganites. We observe that manganite films
grown under biaxial compressive strain exhibit island growth
morphology
which leads to a non-uniform 
distribution of the strain.
Transport and magnetic properties of these
films suggest the coexistence of two different phases,
a metallic ferromagnet and an insulating antiferromagnet.
We suggest that the high strain regions are insulating while the
low strain regions are metallic.
In such non-uniformly strained samples,
we observe a large magnetoresistance and a field-induced
insulator to metal transition. 

\end{abstract}
]
\narrowtext
\tightenlines 

Hole-doped manganites display a remarkable sensitivity to various 
perturbations ~\cite{tok2,tok3}
and such sensitivity results in drastic changes in the sample
properties depending on the form of the sample 
~\cite{mr,mahimr}. 
Thin manganite films display properties different from those of bulk
materials, and several papers have argued that the difference is due to
strain induced by lattice mismatch ~\cite{prellier}.
Lattice mismatch strain is
a biaxial strain which modifies the lattice parameters of the film
and it has been shown that biaxial strain has
an effect which is fundamentally different from that of bulk strain 
~\cite{millis}.
Compressive bulk strain drives the lattice towards cubic symmetry whereas
compressive biaxial strain further distorts the lattice.
It is essential to understand the effect of substrate induced strain
on the manganite thin films to explain the behavior of the thin films
and multilayers of these materials.

In this paper we present
evidence 
that thin films of La$_{0.67}$Ca$_{0.33}$MnO$_3$
($\sim$ 150 \AA~ in thickness)
grown under compressive lattice mismatch strain are structurally,
magnetically and electronically non-uniform. We show that this is due to
structural non-uniformity caused by the island growth mechanism. This 
phenomenon is well established in semiconductor heterostructures-
which are also 
thin films grown under biaxial strain- and has been studied extensively, both
experimentally and theoretically 
~\cite{madhukar,mo,chen,tersoff,priester}.
Studies on the kinetics of the growth of these heterostructures
have shown that the
minimum energy configuration is a non-uniform strain distribution in the
film resulting from a formation of islands. 
A continuous 
wetting layer of a few monolayer thickness covers the substrate first
and islands are nucleated above this layer on further growth of the
film.
This island growth mode leads to a variation in the strain on the film,
both in the direction normal to the substrate and also along the 
plane of the substrate, creating
regions in the film which are strain relaxed (near the top of the
islands) and also some regions (near the periphery of the islands) which have
extremely high strain, i.e. much higher than the lattice mismatch
strain ~\cite{chen}. This type of strain distribution limits the
lateral growth of the islands resulting in uniform island size in the
entire film. This happens due to the 
diffusion of
adatoms away from regions of higher strain as has been observed
experimentally ~\cite{madhukar}. 
All these
factors can lead to structural transitions in the highly strained
regions of the film due to the strain itself and/or due to the
resultant migration of adatoms ~\cite{chen}.

Motivated by the idea that the high sensitivity of the properties of
manganites to changes in structure
and stoichiometry should result in interesting effects when these 
materials are subjected to 
a large non-uniform strain
we have grown thin 
films of La$_{0.67}$Ca$_{0.33}$MnO$_3$ ($\sim$ 150 \AA) with
different amounts of lattice mismatch with the substrate
and have studied the resulting differences in the
growth morphology, magnetization and transport. Conductivity and
magnetization measurements indicate that the film grown under 
compressive strain due to lattice mismatch is a mixture of ferromagnetic
(metallic) and antiferromagnetic (insulating) regions. Atomic
Force Microscopy (AFM) and Transmission Electron Microscopy (TEM)
experiments
confirm the island growth of the strained film and a non-uniform 
distribution of strain over the film. We suggest that the high strain
regions are at the edges of the islands and are insulating and the low-strain
regions are at the top of the islands and are metallic. 
The difference in properties may be
either a direct effect of the strain on the electronic properties
or due to strain induced cation diffusion.

Thin films of La$_{0.67}$Ca$_{0.33}$MnO$_3$ (LCMO), 150 \AA~ in thickness, 
were grown on (001) LaAlO$_3$ (LAO) and (110)
NdGaO$_3$ (NGO) substrates 
by pulsed laser deposition (PLD). 
On LAO there is a
compressive lattice mismatch strain of $\sim$2\% for a film of LCMO
while on NGO this strain
is negligible. The films were grown at a rate of
$\sim$ 1 \AA/sec. The substrate temperature was 820$^{\circ}$C. The
films were grown in an oxygen atmosphere of 400 mTorr. 
The thicknesses were measured by
Dektak IIA profilometer. 
The resistivities were measured by the conventional four-probe
method and the DC magnetization was measured using a SQUID magnetometer.
The lattice parameters were measured using a Siemens D5000 diffractometer
\begin{figure}
\centerline{
\psfig{figure=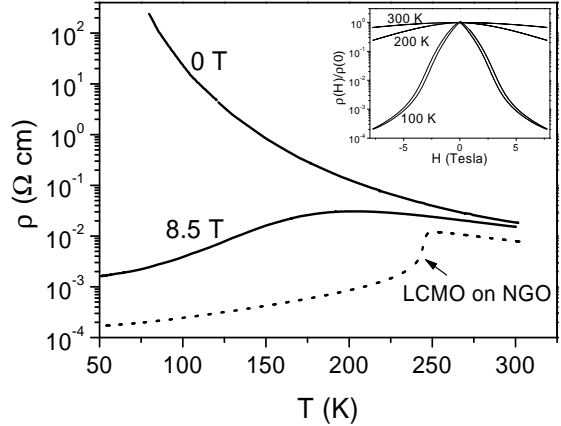,width=8.0cm,height=7.0cm,clip=}
}
\caption{The $\rho$ vs. $T$ behavior of a 150 \AA~
film of LCMO on LAO at 0 T
and 8.5 T (solid lines).
The dotted line
shows the $\rho$ vs. $T$ behavior of a 150 \AA~ film of LCMO on NGO at
0 T. All data were taken while cooling.
The inset shows the normalized $\rho$ vs. $H$ behavior of the film of LCMO on
LAO, at three different temperatures.}
\end{figure}
equipped with a four circle goniometer. 
The in-plane lattice constant measurements showed that
the films were pseudomorphic with the substrate for this range of film
thickness. The nanostructure of the films were measured using a
Nanoscope III AFM operated in the tapping mode. 
Cross section high resolution transmission
electron microscopy measurements were done using a JEOL 4000 EX microscope.

Figure 1 shows the resistivity behavior of a 150 \AA~ film of LCMO on
LAO. The figure also shows the resistivity behavior of a 
150 \AA~ film of LCMO on NGO (dashed line). 
This figure clearly shows the drastic effect
of lattice mismatch strain on the transport properties.
The film of LCMO on LAO 
is insulating. 
In contrast the film of the same thickness
on the lattice matched substrate NGO 
shows a resistivity behavior very close to that of
the bulk.

Figure 2 shows the magnetization of the 
strained film grown on LAO as a function of temperature. 
The magnetization (M) starts rising around 250 K but this rise is much 
slower than what is observed in thicker films of LCMO on LAO
~\cite{jaime}. The inset
shows the $M$ vs. $H$ curve for the film on LAO 
at 5 K. The saturation value of 
$M$ ($M_{sat}$) is $\sim$ 1.8 $\mu_{B}$ which is about 50 \% of the expected
$M_{sat}$ = 3.67 $\mu_{B}$ 
for this compound. This shows that about 
half the volume of the film is not ferromagnetic at low temperatures. The 
magnetization of the film on NGO could not be measured due to the paramagnetic
nature of the substrate, however, 
the correspondence of the resistivity behavior
to that of the bulk compound suggests that the magnetization will be
the same as the bulk material.

A striking feature of our data is that application of a
strong magnetic field causes the low temperature 
\begin{figure}
\centerline{
\psfig{figure=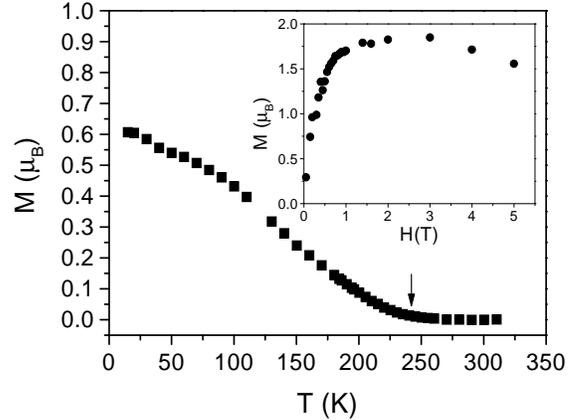,width=8.0cm,height=6.5cm,clip=}
}
\caption{The $M$ vs. $T$ behavior of the 150 \AA~ film of LCMO on LAO in
a field of 1000 Gauss.
The arrow marks the temperature region where the magnetization starts
increasing gradually. The inset shows the $M$ vs. $H$ behavior of the film
at 5 K. The value of $M_{sat}$ is $\sim 1.8 \mu_B$.}
\end{figure}
insulating 
state to become metallic in the strained film.
In figure 1 we show that the resistivity of the LCMO film on LAO in a
field of 8.5 Tesla has an insulator to metal transition near 200 K.
The inset in figure 1 shows the $\rho$ vs. $H$ behavior of the film 
on LAO at 
three temperatures. At
100 K, $\rho$ drops by about 4 orders of magnitude in a field of 8 T.
There is also a significant hysteresis in the $\rho$ vs. $H$ curve at 100 K.
This behavior of the $\rho$ with $H$ is both quantitatively and
qualitatively different from the magnetoresistance behavior
of bulk ceramic La$_{0.67}$Ca$_{0.33}$MnO$_3$. In the latter
the presence of grain boundaries causes
a small but sharp drop in the resistivity at low fields
followed by a gradual decrease in the resistivity at
higher fields ~\cite{mr,mahimr}. Our $\rho$ vs. $H$ data
is also different
from the low field magnetoresistance 
observed in strained ultra-thin films of 
Pr$_{0.67}$Sr$_{0.33}$MnO$_3$ which was attributed to 
domain wall scattering ~\cite{li}. 
On the other hand, the magnitude of the
magnetoresistance and the hysteresis in the $\rho$ vs. $H$ curve at low
temperatures are similar to that observed in materials which exhibit charge
ordering ~\cite{tok1}. 
Our data also resembles that seen in the compound
(La,Pr,Ca)MnO$_3$, now believed to consist of a two-phase
coexistence of ferromagnetic metallic and charge-ordered insulating
phases ~\cite{cheong}, where the field driven insulator to metal transition
is induced by a change of the metal volume fraction through a percolation
threshold. On the basis of this similarity and the magnetic and transport
data discussed above, we argue that the 
biaxially strained thin film of LCMO grown on LAO
exhibits two phase coexistence whereas the film grown on the 
lattice matched substrate NGO does not. 
The origin of the large magnetoresistance in the strained
film is due to this phase separation- the magnetic field drives the 
insulating phase to a metallic phase leading 
to a metallic conduction
path in the film.
We emphasize that for this composition of
LCMO ($x$=0.3), a highly 
\begin{figure}
\centerline{
\psfig{figure=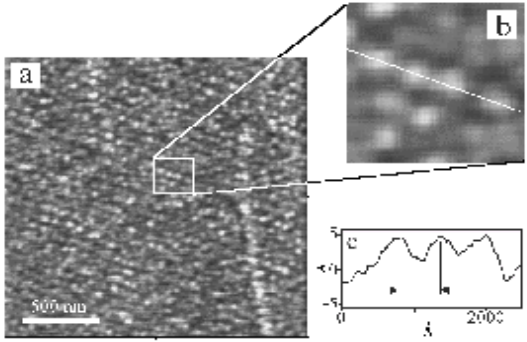,width=8.0cm,height=5.6cm,clip=}
}
\caption{(a) A 2 $\mu$m $\times$ 2 $\mu$m AFM image of the 150 \AA~ film
of LCMO on LAO showing the formation of islands on the film. (b) An enlarged
portion of the image. The profile of the image along the white line is
shown in (c). The distance shown by the two arrowheads in (c) shows the
typical distance between two islands.}
\end{figure}
insulating
state due to phase separation
has not been observed in the bulk form.  

So the important question is: What is the origin of 
the two phase coexistence observed in the film grown under biaxial strain
which results in properties far removed from the properties
of the bulk form of this compound?
To answer this we take AFM images of our films as shown in figures 3 and 4.
At the outset we stress that discontinuities in the film are not the cause.
From the AFM micrographs, we have calculated the roughness of the 
film on LAO to
be about 15 \AA~ which is much smaller than the thickness of the film.
This and the fact that a magnetic field drives the film metallic at low
temperatures, show that the film on LAO 
is continuous. 
It is clear from figures 3 and 4 that there are significant
differences in the nanostructure depending on the strain.
The film of LCMO on NGO has negligible substrate induced strain
and the film shows a step flow growth mode. 
The height of each step is marked in figure 4c. 
The $\rho$ vs. $T$ of this
film is very close to that of bulk LCMO, as shown earlier.
The film of LCMO on LAO which
has about 2\% strain, has an island growth mode.
As shown in figure 1, this film is insulating down to the lowest
temperatures. 
As discussed earlier, island growth leads to a highly
non-uniform distribution of the strain. Such a variation of strain
in the film may also
lead to the migration of the constituent atoms resulting in a
compositional inhomogeneity on the scale of the variation of the strain.
The large strain at the edge of the islands leads to insulating,
and perhaps charge ordered regions due to structural and/or
compositional variations. 
As mentioned earlier the top of the islands are relatively strain free
and these regions are ferromagnetic (metallic)
but are separated by the insulating
regions at the periphery of the islands. Our resistivity data in a field
of 8.5 T suggests that enough 
of the insulating regions are driven metallic at this
field that a metallic path is formed in the sample joining 
\begin{figure}
\centerline{
\psfig{figure=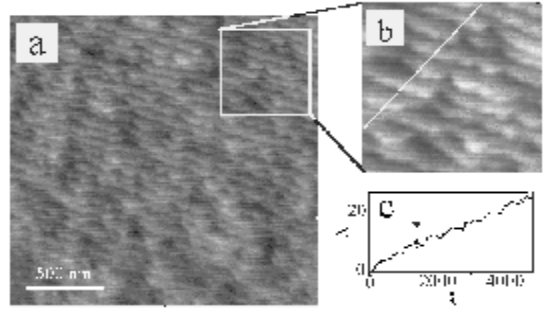,width=8.0cm,height=5.4cm,clip=}
}
\caption{(a) A 2 $\mu$m $\times$ 2 $\mu$m AFM image of the 150 \AA~ film
of LCMO on NGO showing the step flow type growth mode. (b) An enlarged
portion of the image. The profile of the image along the white line is
shown in (c). The two arrowheads define an atomic step of $\sim$ 4\AA.}
\end{figure}
the ferromagnetic
metallic regions in the film and
consequently there is a large drop in the
resistivity of the sample upon the application of a magnetic field. 

The magnetization measurements at 5 K
show that the film on LAO has a saturation magnetization value
of $\sim$ 1.8 $\mu_B$
and the expected $M_{sat}$ for this composition 
is 3.67$\mu_B$ which suggests
that a significant part ($\sim$ 50 \%) of the film is not in the
ferromagnetic state at low temperatures.
The saturation magnetization approaches 3.67 $\mu_B$
as the thickness of the films on LAO is increased i.e,
the effect of the substrate induced strain becomes less.
Another observation is that
on annealing in flowing oxygen at
a temperature of 850$^{\circ}$ for 10 hours, the film on LAO
has the resistivity behavior and lattice
parameters found in thicker films
of LCMO on LAO and a saturation magnetization of 3.4 $\mu_{B}$.
The
AFM images suggest a significant increase in the size of the islands but
more controlled experiments are required. This strengthens our
claim that the insulating behavior is due to strain induced
structural and compositional variations which are removed by annealing the
film in oxygen. 

To get a better picture of the variation of the strain and composition
over the film on LAO, cross sectional TEM (XTEM)
studies were performed on a 1500 \AA~
film of LCMO on LAO. A thicker film was used for reasons of sample 
preparation for the XTEM studies. We assume that the first 150 \AA~ of
this sample is the same as the 150 \AA~ film on LAO.
Figure 5a shows that on the scale of the distance between two
islands (as estimated from the AFM images in figure 3c) 
there is a significant variation in the contrast of the image.
The arrows mark the regions where there is a clear demarcation between two
regions of similar contrast. These are the edges of the islands and 
the distance between the regions marked by the arrows is of the order
of the distance between islands as seen from the profile of the AFM image
shown in figure 3c 
(i.e. $\sim$ 500 \AA). Figure 5b is a 
schematic diagram showing the expected regions of low and high strains. The
variation in the contrast shows a variation of strain and/or 
stoichiometry
in the film, both of which are expected in the growth of these thin
films on lattice mismatched substrates. 
The properties of hole-doped manganites are sensitive to both these factors.
The structure and the stoichiometry affect the transport of the material by
tuning the number and mobility of carriers and the bandwidth.
\begin{figure}
\centerline{
\psfig{figure=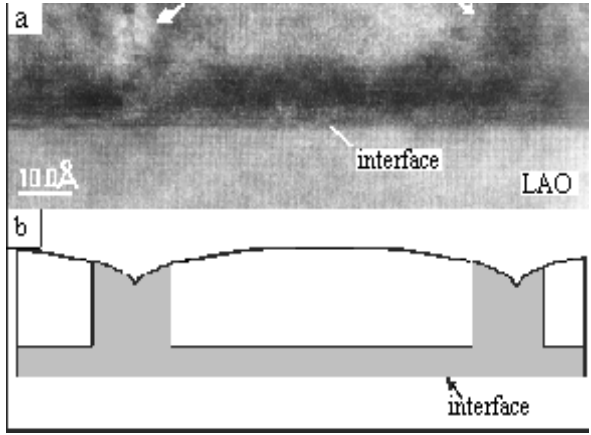,width=8.0cm,height=6.0cm,clip=}
}
\caption{(a) A high resolution XTEM image showing the initial
layers ($\sim$ 200 \AA) of a 1500 \AA~ film of LCMO on
LAO. The white arrows show the edges of the islands where there is a large
variation in the image contrast showing a variation in the strain
and/or stoichiometry. (b) A schematic representation of the regions
of low strain (white regions)
separated by regions of high strain (shaded gray regions).}
\end{figure}
Although very large changes in stoichiometry
are required for producing an effective Ca doping
of $x < 0.2$ or $x > 0.5$, this gives us a possible mechanism 
for having charge ordered (insulating)
regions in the film corresponding to the regions
of very high strain i.e. at the edges of the islands. There is also
a significant variation of the contrast in the image very near the
substrate which reveals the initial wetting layer of the film.
An earlier study of the near-interface transport properties of
La$_{0.67}$Ca$_{0.33}$MnO$_3$ ultra-thin 
films grown on LAO and NGO substrates revealed a surface and interface
related ``dead layer" of about 30 - 50 \AA~ depending on the substrate 
~\cite{sun}. This ``dead layer" could arise due to this wetting
layer. 
We would like to add here that
the effect of tensile strain on the magnetic and transport
properties of LCMO is similar to what is
observed here. Zandbergen et al. ~\cite{zandbergen}
observe a reduced saturation
magnetization and a large magnetoresistance at low temperatures in
their ultra-thin films of LCMO grown on SrTiO$_3$ (STO). 
These properties are 
attributed to the distortions induced in the film due to the lattice
mismatch as inferred from high resolution XTEM experiments. These films,
grown under tensile strain, remain insulating at low temperatures even
upon application of a field of 8 T. In a recent paper F\"{a}th et al. 
~\cite{fath} have
shown scanning tunneling spectroscopy data on thin films of LCMO grown on
STO substrates which suggests a two phase behavior in the film. On the 
application of a magnetic field the metallic phase grows at the expense of
the insulating phase and the authors show a correspondence between this and
the colossal magnetoresistive properties of the material. An LCMO film grown 
on STO is under tensile biaxial strain and should result in a non-uniform
distribution of the strain. The non-uniformity in the strain is a likely
origin of the observed two-phase behavior based on our results
discussed here.

In conclusion, 
we propose the following model to explain the properties of 
LCMO grown on LAO, a film which is under compressive biaxial strain.
The film grows in the form of islands.
The edges of the islands are regions of
high strain and are insulating due to changes in structure and/or 
stoichiometry.
The top of the islands are relatively strain free and only these parts
are ferromagnetic and conducting at low temperatures
and thus a two-phase state is formed. This explains the 
reduced saturation magnetization at low temperatures. The insulating
regions separating the islands makes the film insulating, down to the 
lowest temperatures. The insulating regions are driven to a metallic
state upon application of a magnetic field which results in a
large decrease in the resistivity of the film. For a direct measure of the 
magnetization in different parts of the film low temperature magnetic force
microscopy measurements in the presence of a magnetic field are underway.

This work is partially supported by
the MRSEC program of the NSF at the
University of Maryland, College Park (Grant DMR96-32521). AJM acknowledges
NSF-DMR-9705482.


\end{document}